# Quantifying the AI Gap: A Comparative Index of Development in the United States and Chinese Regions

Yuanxi Li[1], Lei Yin[2*]

[1]Department of Science and Technology Studies University College London, United Kingdom

[2]Meta-Fusion Technology Co., Ltd

richard.li.24@ucl.ac.uk, yinlei@metafusion.cc

**Abstract** This study develops a comprehensive Artificial Intelligence (AI) Index with seven primary dimensions, designed for provincial-level and industry-specific analysis. We employ an anchor point method for data normalization, using fixed upper and lower bounds as benchmarks, and devise a hierarchical indicator weighting system that combines expert judgment with objective data. The index draws from authoritative data sources across domains including official statistics, patents and research outputs, education and talent, industrial economy, policy and governance, and social impact. The China-US comparison indicates that under a unified framework, the US composite score (68.1) exceeds China's (59.4). We further dissect China into seven main areas to form a sub-national index. The findings reveal stark regional disparities in China's AI development: the North, East, and South regions lead in composite scores, whereas central and western regions lag significantly, underscoring the effects of regional concentration of innovation and industry resources. This research provides an academic reference and decision support tool for government agencies and research institutions, informing more targeted regional AI development strategies.

## 1. Introduction

The competition in Artificial Intelligence between the United States and China defines the current technological era, yet a precise, multi-dimensional quantification of the "AI Gap" remains elusive. Most existing indices focus on national-level readiness or track individual metrics without aggregating them into a unified score for direct comparison. This makes it difficult for policymakers and stakeholders to gauge overall progress and identify specific areas of strength and weakness.

To address these gaps, this study introduces a novel methodology to benchmark this gap, finding that the **US leads China with a composite AI Index score of 68.1 to 59.4**.

---

[*] Corresponding author

Our analysis dissects this 8.7-point gap across seven key dimensions, revealing that the US maintains a significant lead in **Policy & Governance** and **Industry & Economy**, while China surprisingly outperforms in **Social Impact**.

Furthermore, we extend this analysis to China's diverse sub-national regions for the first time, mapping the country's internal AI landscape. The results quantify a stark divide between a few highly developed coastal hubs (e.g., Shanghai and Beijing regions) and the lagging inland provinces. This research provides a quantitative tool for benchmarking AI development, positioning the performance of Chinese regions in a global context and helping to identify areas of leadership and weakness relative to a world-leading benchmark.

2. Methodology: A Multi-dimensional Index

To construct the index, we developed a comprehensive assessment framework based on an anchor method and a layered weighting approach.

**2.1. Normalization and Weighting** Unlike traditional Min-Max normalization, which is dependent on the samples in a dataset, we use an **anchor method**. Each indicator is normalized using predetermined high and low anchor values (e.g., international benchmarks or policy goals). This ensures that scores remain stable and comparable across regions and over time, even as new data is added.

A hierarchical indicator weight tree was designed to reflect the relative importance of each domain to overall AI capability. The seven primary dimensions and their corresponding weights are detailed in Table 1. This mixed weighting strategy combines expert judgment with data-driven insights, emphasizing critical domains like R&D and Industry while ensuring no single dimension dominates the score.

Table 1: AI Index Dimensions and Weights

| Dimension | Weight | Rationale for Weighting |
|---|---|---|
| Research & Development | 20% | Represents the fundamental engine for innovation and long-term potential. |
| Industry & Economy | 20% | Measures the scale of the AI market, investment, and industrial application. |
| Technical Performance | 15% | Assesses cutting-edge capabilities, such as frontier AI models. |

| Dimension | Weight | Rationale for Weighting |
|---|---|---|
| Education & Talent | 15% | Evaluates the human capital pipeline, from graduates to top-tier experts. |
| AI for Science | 10% | Gauges the application of AI in advancing fundamental scientific research. |
| Policy & Governance | 10% | Reflects the institutional environment, including regulations and strategic planning. |
| Social Impact | 10% | Measures the societal adoption and penetration of AI technologies in daily life. |

**2.2. Data Sources** The data for our index are drawn from diverse and authoritative sources to comprehensively cover all aspects of AI development. Main sources include:

- **Official Statistics:** National Bureau of Statistics (China) data on regional macroeconomics, R&D expenditure, and graduates in relevant majors.

- **Patents and Innovation:** Data from the World Intellectual Property Organization (WIPO) and the China National Intellectual Property Administration (CNIPA). According to 2023 statistics, China accounted for 69.7% of global AI patents, far surpassing other nations [1].

- **Research Publications:** Data from Microsoft Academic, Google Scholar, OpenAI, Tsinghua's AMiner, and the Stanford AI Index to track publication output and conference papers.

- **Investment and Industry:** Venture investment data from Crunchbase and IT Juzi, along with industry reports on AI market scale and industrial robot installations.

- **Education and Talent:** Ministry of Education data, professional networks (e.g., LinkedIn), and Tsinghua University's "China AI Talent Report."

- **Policy and Governance:** Sub-indicators from the Oxford Insights Government AI Readiness Index and counts of national and provincial policy documents.

- **Societal Adoption and Impact:** Survey data from the Ministry of Industry and Information Technology (MIIT) on AI adoption rates and scraped social media and news data.

## 3. Results and Analysis

**3.1. The China-US AI Gap** According to our calculations, China's composite AI index score is **59.4**, while the US scores **68.1**, resulting in an 8.7-point gap. This indicates that the overall AI development level of the US is moderately higher than China's. Figure 1 provides a side-by-side comparison across the seven dimensions, illustrating the structural causes behind the overall gap.

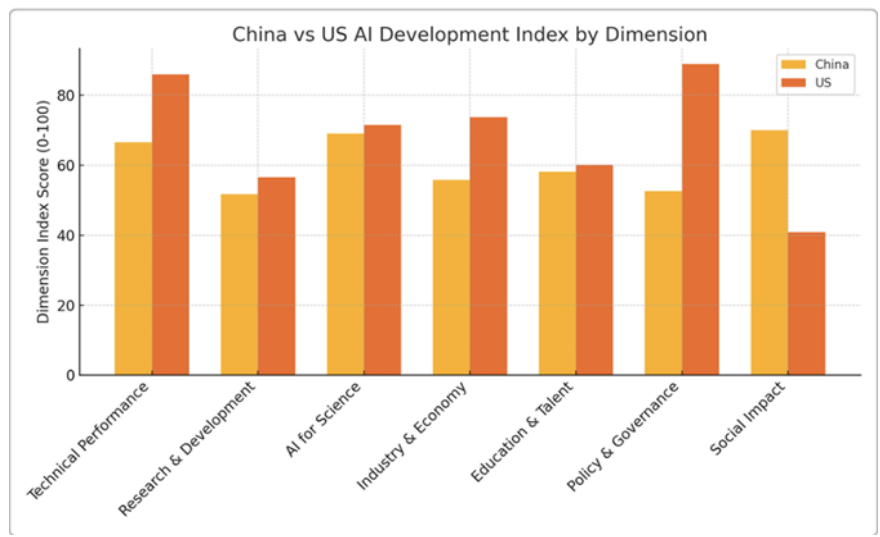

Fig.1 China vs US AI Index by Dimension

A dimensional breakdown reveals the story behind the numbers:

- **Largest Gaps: Where the US Leads.** The US holds its most significant advantages in **Policy & Governance** (8.9 vs. 5.3) and **Industry & Economy** (14.8 vs. 11.2). The US governance framework is more mature, with federal and state actions on AI legislation increasing dramatically in recent years [2], whereas China has room to improve in terms of comprehensive regulations and regulatory quality [3]. The industry gap is driven by a vastly larger investment ecosystem; Stanford's report showed that in 2024, US private AI investment was nearly 12 times China's [4].

- **Moderate Gaps.** The US also leads in **Technical Performance** (12.9 vs. 9.98), reflecting its edge in frontier model capabilities. The Stanford AI Index notes that in 2024, 40 notable AI models came from US-based institutions versus 15 from China [5]. The gap in **Research & Development** (11.3 vs. 10.3) is narrowing, as China has surpassed the US in the sheer number of AI research publications, but the US still leads in highly cited papers and fundamental research breakthroughs [6].

- **The Surprise: Where China Leads.** China outperforms the US in one

dimension: **Social Impact** (7.0 vs. 4.1). This reflects a higher degree of societal acceptance and penetration of AI. AI-driven applications in mobile payments and smart cities are extremely widespread, and China's use of industrial robots is the highest in the world by volume, accounting for 51.1% of global industrial robot installations in 2023. [7]

- **Near Parity.** The two countries are very close in **AI for Science** (7.15 vs. 6.91) and **Education & Talent** (9.03 vs. 8.72). While the US has an edge in top-tier AI talent and its ability to attract global expertise, China has a massive pipeline of AI talent and is rapidly expanding its talent pool.

**3.2. Sub-national Disparities in China's AI Development** The national score of 59.4 masks extreme regional imbalances within China. Our sub-national index reveals a clear "core-periphery" gap, with three coastal regions far ahead of the rest of the country.

Tab. 1 : AI Index by Chinese Regions

| Region | Technical Performance | Research & Development | AI for Science | Industry & Economy | Education & Talent | Policy & Governance | Societal Impact | Composite Index |
|---|---|---|---|---|---|---|---|---|
| East China (Shanghai & East) | 71.7 | 27.5 | 31.7 | 35.5 | 31.7 | 38.5 | 48.8 | **40.0** |
| North China (Beijing & North) | 64.3 | 28.5 | 37.3 | 24.6 | 28.4 | 39.1 | 37.0 | **35.9** |
| South China (Guangdong) | 62.9 | 19.3 | 14.3 | 20.8 | 27.8 | 25.1 | 28.2 | **28.4** |
| Southwest China (Sichuan/Chongqing) | 27.3 | 7.0 | 3.3 | 6.8 | 2.7 | 13.0 | 12.3 | **10.1** |
| Central China (Hubei/Hunan) | 25.4 | 7.7 | 6.5 | 6.8 | 4.2 | 11.9 | 16.3 | **10.8** |
| Northwest China (Shaanxi) | 26.7 | 6.2 | 4.0 | 3.3 | 2.9 | 9.4 | 9.4 | **8.6** |

| Region | Technical Performance | Research & Development | AI for Science | Industry & Economy | Education & Talent | Policy & Governance | Societal Impact | Composite Index |
|---|---|---|---|---|---|---|---|---|
| Northeast China (Liaoning etc) | 37.6 | 3.9 | 2.9 | 2.2 | 2.3 | 6.8 | 7.3 | **8.9** |

- **Tier 1: The Leaders.** East China (led by Shanghai) and North China (led by Beijing) are in a close race for the top spot, with scores of **40.0** and **35.9**, respectively. East China's lead is driven by its strength in Industry & Economy and Societal Impact, reflecting its powerful manufacturing and internet base. North China remains the undisputed hub for R&D and Policy & Governance.

- **Tier 2: The Strong Follower.** South China (led by Guangdong/Shenzhen) ranks a clear third with a score of **28.4**. Its strengths are a vibrant tech sector (e.g., Tencent, Huawei) and a strong manufacturing base, but it lags the top two in research and governance depth.

- **Tier 3: The Lagging Regions.** The Central, Southwest, Northwest, and Northeast regions all score in the low single digits to low teens (ranging from 8.6 to 10.8). This quantifies the often-cited regional disparity in China's tech landscape [8]. These four regions significantly trail the coastal hubs, contributing little in most dimensions and highlighting a deep regional disparity in China's AI landscape.

## 4. Conclusion

This study introduced an AI index that quantifies the capabilities of China and the US on a unified scale. We find that the US's overall AI index is about 8.7 points higher than China's (68.1 vs. 59.4), indicating a moderate but significant lead. This gap is primarily driven by US advantages in the **Policy & Governance** and **Industry & Economy** dimensions; the US far outstrips China in AI institutional environment and investment scale [9]. In contrast, China performs better in **Social Impact**, with a higher penetration of AI applications in its society. The differences in R&D, talent, and technical performance are relatively small and narrowing [10].

At the sub-national level, AI development in China is highly imbalanced. A few eastern regions (notably the Yangtze River Delta, the Beijing area, and the Pearl River Delta) are driving the vast majority of progress, while vast inland regions lag far behind. Encouragingly, some cities in these lagging regions, like Chengdu, Xi'an, and Hefei, are emerging as new AI hubs in their own right [11].

While this study provides meaningful findings, we recognize its limitations. Data availability for some sub-national indicators remains suboptimal, requiring reliance on estimates. Moreover, the rapidly evolving nature of AI, particularly the surge in foundation models in 2024, is not fully captured in our current index. Future work should establish more real-time data updating mechanisms and potentially extend this comparative methodology to other regions, such as the EU, to build a global regional AI index.